\documentclass[rnote,traditabstract]{aa} 
\usepackage{graphicx}
\usepackage{txfonts}
\begin{document}
   \title{An educated search for transiting habitable planets:} 
   \subtitle{Targetting M dwarfs with known transiting planets}
      
   \author{M. Gillon
          \inst{1, 2}, X. Bonfils\inst{2,3}, B.-O. Demory\inst{4,2}, S. Seager\inst{4}, D. Deming\inst{5}, A.~H.~M.~J. Triaud\inst{2}
          }

   \institute{Institut d'Astrophysique et de G\'eophysique,  Universit\'e
  de Li\`ege,  All\'ee du 6 Ao\^ut 17,  Bat.  B5C, 4000 Li\`ege, Belgium \\
                \email{michael.gillon@ulg.ac.be}
         \and
             Observatoire de Gen\`eve, Universit\'e de Gen\`eve, 51 Chemin des Maillettes, 
             1290 Sauverny, Switzerland
          \and   
             Universit\'e Joseph Fourier Ð Grenoble 1, Centre national de la recherche scientifique, 
             Laboratoire d'Astrophysique de Grenoble (LAOG), UMR 5571, 38041 Grenoble Cedex 09, France
          \and         
           Department of Earth, Atmospheric and Planetary Sciences, Department of Physics, Massachusetts 
           Institute of Technology, 77 Massachusetts Avenue, Cambridge, MA 02139, USA
                     \and
           Planetary Systems Branch, Code 693, NASA/Goddard Space Flight Center Greenbelt, MD 20771, USA\\ 
       }

   \date{Received; accepted}
 
       \abstract{Because the planets of a system form in a flattened disk, they are expected to 
share similar orbital inclinations at the end of their formation. The high-precision photometric 
   monitoring of stars known to host a transiting planet could thus reveal the transits of one or 
          more other planets. We investigate here the potential of this approach for the M dwarf 
GJ~1214 that hosts a transiting super-Earth. For this system, we infer the transit probabilities 
as a function of orbital periods. Using Monte-Carlo simulations we address both the cases for 
fully coplanar and for non-coplanar orbits, with three different choices of inclinations distribution
 for the non-coplanar case.  GJ~1214 reveals to be a very promising target for the considered approach. 
 Because of its small size, a ground-based photometric monitoring of 
     this star could detect the transit of a habitable planet as small as the Earth, while a 
 space-based monitoring could detect any transiting habitable planet down to the size of Mars. 
 The mass measurement of such a small planet would be out of reach for current facilities, but 
 we emphasize that a planet mass would not be needed to confirm the planetary nature of the transiting object. 
 Furthermore, the radius measurement combined with theoretical arguments 
 would help us to constrain the structure of the planet.
}

\authorrunning{M. Gillon et al.}
  \titlerunning{Targetting M dwarfs with known transiting planets}

   \maketitle
%

\section{Introduction}

One of the most significant goals of modern astronomy is the detection and 
characterization of planets similar to our Earth: a terrestrial planet lying in 
the habitable zone (HZ; e.g., Kasting et al. 1993) of its host star. That will 
place our Earth into context, as just the closest member of the habitable telluric 
planet class, and will possibly establish whether life exists elsewhere in the Universe.
While direct detection of habitable terrestrial planets around solar-type stars 
are envisioned with future ambitious missions like Terrestrial Planet Finder (e.g., Traub et al. 2007) 
and Darwin (e.g., Cockell et al. 2008),  technological developments are still needed 
to make these missions successful, and none of them is fully funded. In this context, the 
indirect detection of terrestrial planets transiting nearby M dwarfs represent a promising 
shortcut (e.g., Charbonneau  2009). Indeed, the planet-to-star contrast for an
Earth-size planet orbiting in the HZ of an M dwarf is much more favorable than for the 
Earth-Sun system, permitting the detection of atmospheric biosignatures by eclipse 
spectroscopy with the planned James Webb Space Telescope (JWST, e.g., Seager et al. 2009, 
Kaltenegger \& Traub 2009) without the challenging need to separate the light of the planet 
from that of its host star. 

 Two different approaches are presently used to detect low-mass planets 
 transiting nearby M dwarfs. 
 
 \begin{enumerate}
 
 \item Doppler surveys targeting nearby M dwarfs have 
 detected several low-mass planets, including a few `hot Neptunes'  and
  `super-Earths'\footnote{A super-Earth is loosely defined as 
 a planet between 1 and 10 Earth masses.}. The subsequent search for the transits of
  these planets revealed the transiting nature of GJ~436b (Gillon et al. 2007b), the 
  first transiting planet significantly smaller than Jupiter. A growing number of habitable 
  super-Earths are expected in the near future and, clearly, this effort should be 
  pursued and  intensified, because only a substantial increase of these detections 
  ($\sim\times$ 50-100) will reveal  transiting habitable super-Earths.

\item Most of the known transiting planets have been detected by dedicated 
photometric surveys monitoring thousands of stars in fairly large fields of view.
Nevertheless, nearby M dwarfs are spread all over the sky, so most transit 
 surveys do not probe enough of them to make a transit detection likely. An 
 alternative approach is used by the MEarth Project (Nutzman \& Charbonneau 2008). 
This survey individually monitors nearby M dwarfs with eight 40cm telescopes 
located at Mt. Hopkins, Arizona. Thanks to the small size of its targets, MEarth 
is sensitive to transiting planets down to a few $R_\oplus$, as demonstrated  
by its recent detection of a 2.68 $R_\oplus$ super-Earth transiting GJ~1214 
(Charbonneau et al. 2009, hereafter C09).  After CoRoT-7b (L\'eger et al. 2009, 
Queloz et al. 2009), GJ~1214b is the second super-Earth caught in transit. 
Unlike CoRoT-7b, GJ~1214b is a good target for an atmospheric 
characterization with, e.g., the future JWST, thanks to the small size and infrared 
brightness of its host star. Nevertheless, the approach used by MEarth is young 
and its effectiveness to detect habitable planets is difficult to assess.
 
 \end{enumerate}

We outline here the potential of a third approach to discover habitable planets
transiting nearby M dwarfs. Its main principle is simple: planets form within disks, 
therefore a planet orbiting in the HZ of a given star should have a higher $a$ $priori$ transit 
probability if its star harbors a known transiting planet. This assumption is  
supported by the small scatter of the orbital inclinations of the eight planets of the solar 
system ($rms$ = 2.2 deg) and of the regular satellites of Jupiter ($rms$ = 0.35 deg) and 
Saturn ($rms$ = 2.8 deg). For dynamical stability reasons, the fact that low-mass planets 
systems (e.g., GJ~581, HD~40307) detected by Doppler surveys tend to be `packed' 
also favors a small scatter of the orbital inclinations. Recent $Kepler$ results also support 
this assumption (Steffen et al. 2010, Holman et al. 2010). Depending on the orbital inclination 
of the known transiting planet, on the assumed distribution of the orbital inclinations 
of the  planetary system, on the size of the star, and on  its physical distance to its HZ, 
significantly enhanced transit probabilities can be expected for habitable planets. A 
dedicated high-precision photometric monitoring of M dwarfs known to harbor close-in 
transiting planets could thus be an efficient way to detect transiting habitable planets in 
the near future. The aim of this Research Note is to assess the potential of this approach 
for the only M dwarf presently known to host a transiting super-Earth, GJ~1214. Section 2 
presents our computational method. Our results are given in Sect. 3 and are discussed in Sect. 4.


\section{Transit probabilities for additional planets}

To account for the uncertainty of the measured orbital inclination for the known planet and to 
include the possibility that the searched planet does not share the exact same inclinations, 
we compute the transit probability with Monte-Carlo simulations.  

We define a  `terrestrial' region extending from twice the period of the transiting planet to the ice line 
defined as (Ogihara \& Ida 2009)

\begin{equation}
r_{ice} = 2.7 \bigg( \frac{L_\ast}{L_\odot}\bigg)^2 \textrm{ AU,}
\end{equation}
where $L_\ast$ and $L_\odot$ are the luminosity of the star and the Sun. For 
GJ~1214, we take $L_\ast = 0.00328$ $L_\odot$ for GJ~1214 (C09). 
This luminosity translates into $r_{ice}$ = 0.155 AU. Using  $M_\ast = 0.157$ $M_\odot$ 
for GJ~1214 (C09),  the corresponding orbital period is only 56 days.

We assume circular orbits for all planets. We divide the deduced terrestrial region into 1000 equal steps
 in semi-major axis. For each semi-major axis $a_i$ ($i=1:1000$), 10,000 orbital inclinations $i_k$ ($k=1:10000$) 
are drawn via

\begin{equation}
i_{k} \sim N(i_{tp},\sigma_{tp}^2+\sigma_{disk}^2) \textrm{ ,}
\end{equation} 
where $N(m,n^2)$ represents the normal distribution of mean $m$ and variance $n^2$, 
$i_{tp}$ and  $\sigma_{tp}$  are the orbital inclination of the known transiting planet and 
its 1-$\sigma$ error, and $\sigma_{disk}$ is the assumed standard deviation of the 
orbital inclinations in the planetary systems. We adopt $i_{tp} =  88.62 \pm 0.35 $ deg 
(C09), and $\sigma_{disk}$ = 2.2 deg, 
the corresponding value for the eight planets of the solar system (Murray \& Dermott 2000). 
We also test values twice smaller and larger for $\sigma_{disk}$, i.e. 1.1 and 4.4 deg, and the 
unrealistic value $\sigma_{disk}$ = 0 deg to illustrate the influence of the inclination scatter 
in the planetary systems. 

For each orbital inclination $i_k$ drawn via Eq.~2, a transit impact parameter $b_k$ is 
computed via

\begin{equation}
b_k = \frac{a_i}{R_\ast} \cos{i_k} \textrm{.}
\end{equation}  We use $R_\ast$ =  0.211 $R_\odot$ 
(C09).  If  the absolute value of $b_k$ is lower than 1, a transit is recognized. If so, 
the transit duration $D_k$  is computed via 
\begin{equation}
D_k = \frac{P_i R_\ast}{\pi a} \sqrt{1- b_k^2} \textrm{,}
\end{equation} 
where $P_i$ is the orbital period. Eq. 4 assumes that the planet has a negligible size 
compared to the star (e.g., Seager \& Mall\'en-Ornelas 2003). 

For each step in semi-major axis, the transit probability is computed as the fraction of the 10,000 
drawn inclinations leading to a  transit. We also compute for each step the geometric transit 
probability $P_{tr}$, neglecting the transiting nature of the known planet, using
\begin{equation}
P_{tr} = \frac{R_\ast}{a} \textrm{.}
\end{equation} 

Following Kasting et al. (1993), we define the inner edge $HZ_{in}$ and outer edge $HZ_{out}$ of 
the habitable zone as 
\begin{eqnarray}
HZ_{in} = 0.95 \bigg( \frac{L_\ast}{L_\odot}\bigg)^2 \textrm{ AU,}\\
HZ_{out} = 1.37 \bigg( \frac{L_\ast}{L_\odot}\bigg)^2 \textrm{ AU.}
\end{eqnarray} From these formula, the HZ of GJ~1214 extends from 0.054  to 0.078 AU.
These edges correspond to orbital periods of 11.6 and 20.1 days. We finally 
average the transit probabilities and durations for the whole HZ to obtain a representative value
for GJ~1214.

\section{Application}

Neglecting the transiting nature of GJ~1214b, a planet in the HZ of GJ~1214 would have a 
mean transit probability of only 1.5\%. Taking into account the transits of GJ~1214b and assuming 
$\sigma_{disk}$ = 2.2 deg leads to a much higher transit probability of 25.1 \% (Fig.~1), the mean
 expected duration of a transit being 1.4 hour. A scatter twice larger for 
 the orbital inclinations in the GJ~1214 planetary system would lead to a reduced transit probability of 
 14.8\%, which is still ten times higher than the probability expected if we do not consider that GJ~1214b  does transit. For $\sigma_{disk}$ = 1.1 deg, the probability goes up to 30.1 \%, while it goes down to 7.7\%  for fully coplanar orbits. The difference between these latter two values well illustrates the advantage given by the fact that planets of the same system are probably not perfectly coplanar.
 
Assuming $\sigma_{disk}$ = 2.2 deg and that a terrestrial planet orbits in the HZ of GJ~1214, a 
constant photometric monitoring of the star during 20.1 days would therefore have a $\sim$ 25 \% probability to catch at least one of its transits, if a sufficiently high photometric precision is reached. Two main options can be considered to perform this photometric search: a multi-site ground-based survey using several telescopes spread in longitude or a space-based monitoring using an instrument able to stare at GJ~1214 for three weeks. 
  
 Ground-based photometric time series reaching the sub-mmag  precision level for a time sampling better than one minute have been  obtained in the optical for several transiting planets (e.g., Gillon et al. 2009, Johnson et al. 2009, Winn et al. 2009). Our own analysis of most of these data makes us conclude that a 
 transit with a depth as low as one mmag could be firmly detected with a such a photometric 
 precision. Shallower occultations of short period transiting planets could be detected from the 
 ground (e.g., Sing \& L\'opez-Morales 2009), but only because the expected timing and shape of the 
 occultations are  known via the analysis of the transits and radial velocities (RV). For a transit that could happen anytime during a run of two months, a detection limit better than one mmag seems unrealistic 
 with current instruments and techniques. Furthermore, the most precise ground-based differential light curves were obtained during nights with excellent atmospheric conditions and at low airmass, far from the average observation conditions of a multi-site survey observing the same star during three weeks. For these reasons, the 
actual detection limit for such a program would probably be closer to two than one mmag. 
  
Figure 1 (middle panel) shows that with this photometric precision, a ground-based monitoring campaign of GJ~1214 would be sensitive to very small planets. Planets as small as the Earth or 
even smaller could be detected. It is worth noticing here that the detection limit achievable from the ground is mostly limited by the atmosphere and the photometric correlated noises that it creates.

The detection limit could be much better for a space-based telescope. For instance, the 
{\it Spitzer} telescope has produced several high-precision photometric time-series for another M-dwarf, GJ~436 (e.g., Gillon et al. 2007a, Deming et al. 2007). {\it Spitzer}'s cryogen is now depleted, but the telescope is  still active (under the name `$Warm$ {\it Spitzer}' ) and keeps its full potential in the two bluest channels (3.6 and 4.5 $\mu$m) of its IRAC camera (Stauffer et al. 2007). Our analysis of the occultation 
photometry obtained for GJ~436b at 3.6 $\mu$m  (Lanotte et al., in prep.) allows us to conclude that 
an eclipse of 200 ppm would be firmly detected for GJ~436 within the 3.6 $\mu$m channel of
{\it Spitzer}. At 3.6 $\mu$m, GJ~1214 is 2.7 magnitudes fainter than GJ~436. Considering a complete noise model and {\it Spitzer} instrumental throughput corrections\footnote{http://ssc.spitzer.caltech.edu/documents/som/}, we obtain a SNR of $\sim$990 for a 12s exposure at 3.6 $\mu$m. This translates into a theoretical error of  450 ppm per minute, and 59 ppm per hour. Considering only white noise, we would therefore conclude 
  that a transit of 240 ppm and lasting one hour could consequently be detected at 4-$\sigma$. Still, fluxes measured 
  with the InSb detectors of IRAC show a strong correlation with the changing position of the star on the array 
  (e.g., Knutson et al. 2008). This purely spatial `pixel-phase' effect is due to the combination of the 
  undersampling of the stellar image, the intra-pixel sensitivity, and the jitter of the telescope. It can be well 
  modeled by an analytic function of the $x$ and $y$ coordinates of the stellar image center, but the inaccuracy  of this model and the finite precision on the measured stellar positions result in a residual photometric correlated  noise with a timescale similar to that of a typical transit. From our experience with GJ~436 (Lanotte et al., in prep.) and CoRoT-2 (Gillon et al. 2010), the impact of the `pixel-phase' effect on the final photometric precision is SNR-dependent, at least in the high SNR regime. Basically, it leads to errors on the fitted parameters (including the eclipse depth) $\sim$twice larger than expected when considering only white noise. In the case of GJ~1214, a transit lasting  one hour would thus need a depth of $\sim$450 ppm 
  to be detected at 4-$\sigma$. This limit is shown in Fig.~2. It corresponds to a planet 
 size of $0.49$ $R_\oplus$, i.e. smaller than Mars (0.53 $R_\oplus$). 
 
 Alternatively, one may await the RV detection of any additional planet {\it before} undertaking the 
photometric search for its transit. Not only would the presence of the planet be known with certainty, 
but the observational window would also be narrowed with an {\it a priori} ephemeris. To estimate the 
observational effort required for a RV detection for GJ~1214 ($V$=15),  we compute the detection limit 
(Zechmeister \& K\"urster 2009) imposed by  the 28 HARPS RVs reported in the detection paper (C09).  
Figure 1 (bottom panel) shows the semi-amplitude above which a planet would have been detected, with 
a 99\% confidence level (for all of our 12 trial phases). The plot shows fluctuations because we have too few data points for a clean sampling, but more data will smooth the curve and it will eventually be independent of the period. More data will scale down the detection limit and, anticipating a better sampling from these additional points, a  K$\sim$10 m.s$^{-1}$ limit (indicated by a dashed line) seems a better estimate for the velocity amplitude we aim to scale. To detect an Earth-mass planet orbiting at 0.066 AU -- or K$\sim$0.88 m.s$^{-1}$ -- we will need 130 times more data points ($> 2000$ h). A true Earth-mass planet (or lower mass) requires therefore unrealistic observing time with current velocimeters.
 
  \begin{figure}
   \centering
   \includegraphics[width=8cm]{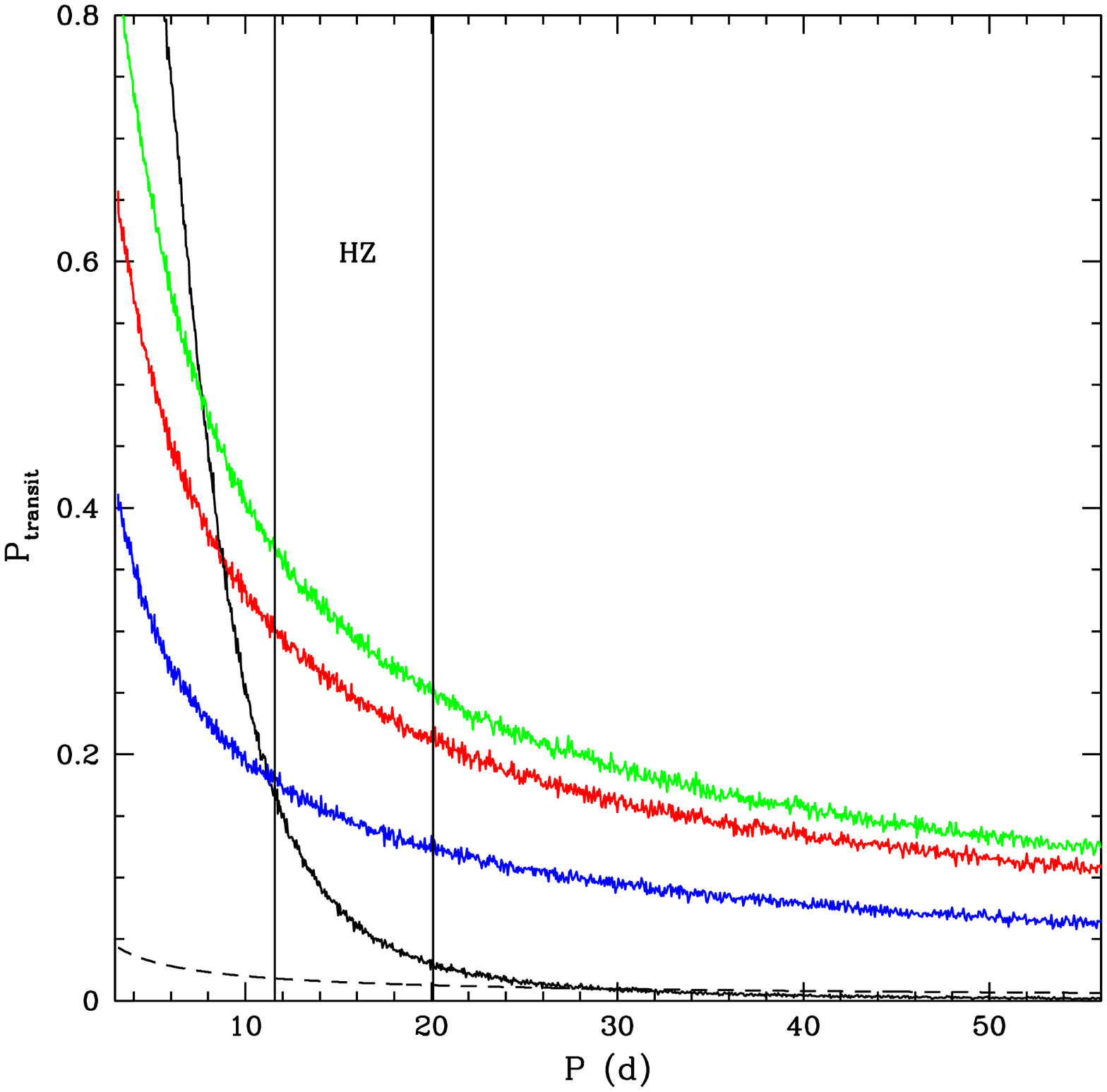}
    \includegraphics[width=8cm]{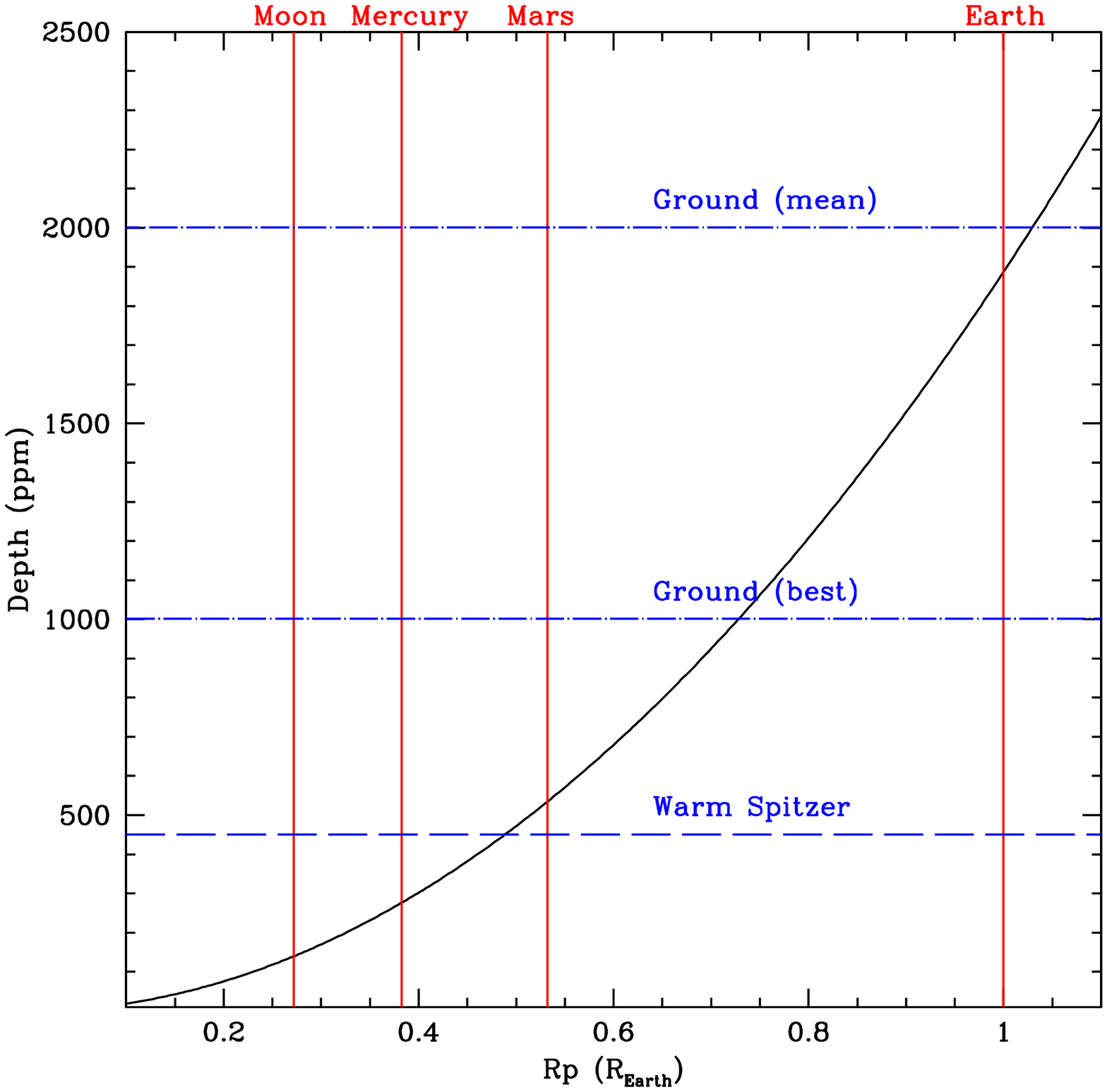}
    \includegraphics[width=8cm]{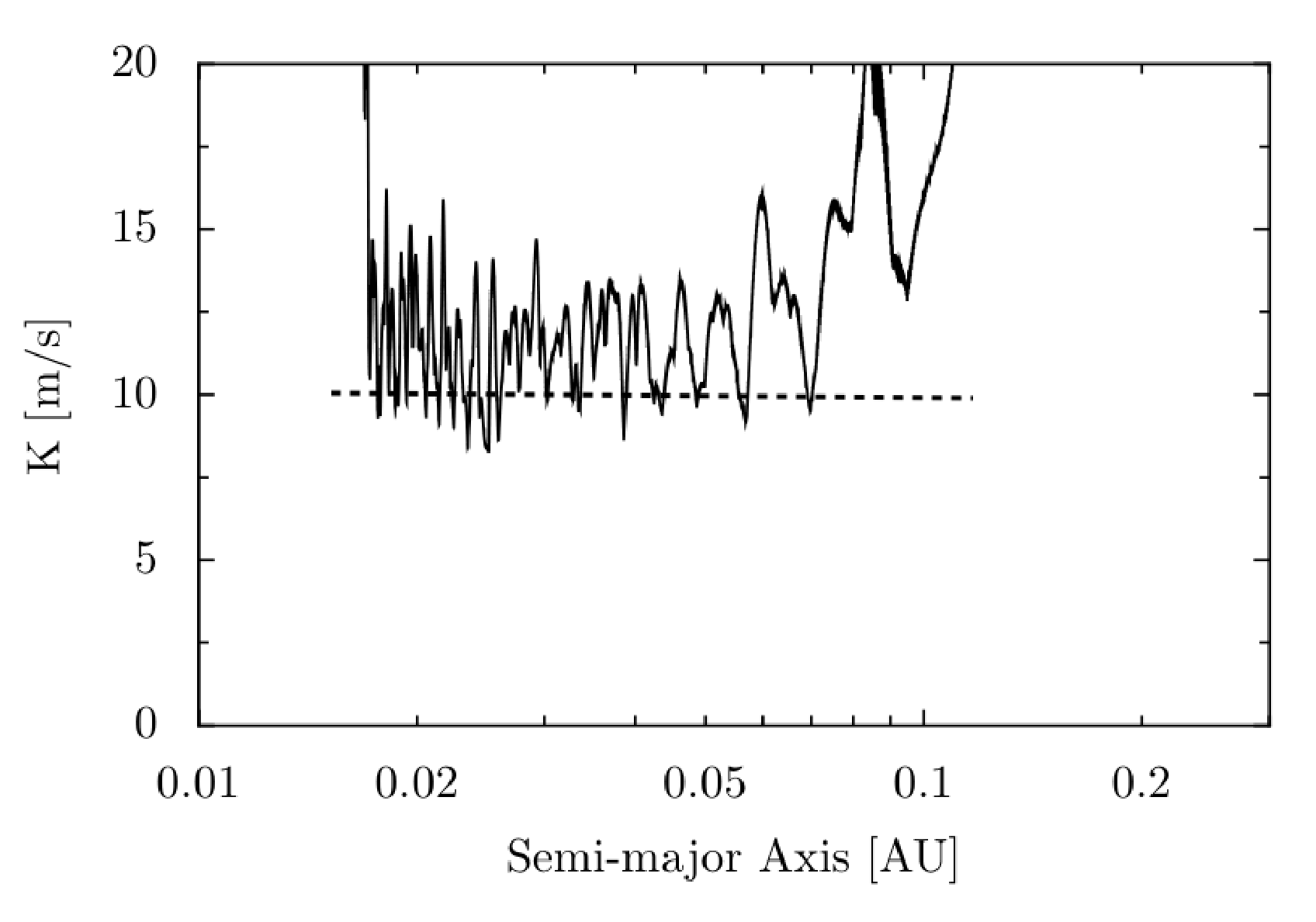}
    \caption{$Top$: transit probability for a planet in a circular orbit around GJ~1214 as a function 
    of the orbital period, assuming $\sigma_{disk}$ = 0 (black), 1.1 (green), 2.2 (red) and 4.4 deg (blue), 
    and neglecting the transiting nature of the known planet (dashed line). 
    $Middle$:  transit depth of a planet orbiting around GJ~1214 as a function of the planetary radius. $Bottom$: 
    RV semi-amplitude above which a planet could be detected in the HARPS data. See text for 
    details. }
     \label{}
   \end{figure}     

  
  \section{Discussion}
 
GJ~1214 appears to be a promising target for the approach described here. Because of its 
small size and luminosity, the transit probability in its HZ is fairly large, and very small planets could 
be detected transiting this star. Another major advantage of GJ~1214  is the proximity of its HZ. To probe its HZ, a constant monitoring of the star during only three weeks would be needed.
For planets orbiting in the inner part of the HZ, two transits could be observed during such a run of three weeks, leading to an improved sensitivity to very small planets. 

For the area interior to the HZ, the transit probability goes up to 
a mean value of 44 \% for GJ~1214 (asssuming $\sigma_{disk}$ = 2.2 deg). Because of the larger 
number of transits observed for shorter periods, the sensitivity to smaller planets would be better 
than for the HZ. For instance, $Warm$ {\it Spitzer} could then detect a planet as small as Mercury (0.38 $R_\oplus$).

Discovery of a transiting habitable-zone planet as small as Mars would create a very challenging mass measurement for the RV method. Mars is nearly ten times less massive that the Earth, and the mean semi-amplitude of the RV wobble due to a Mars-mass planet orbiting in the HZ of GJ 1214 would be only 9.4 cm.s$^{-1}$, while we have seen in Sect. 3 that an Earth-mass planet producing a RV signal ten times larger would be out of reach for current spectrographs. Of course, with the {\it a priori} knowledge of the phase and period, one would need to sample the RV orbit at its extrema only, facilitating the mass measurement of an Earth-mass planet. Still, the measure seems difficult and for sure, measuring the mass for a Mars-mass planet would remain  out of reach for current instruments. To measure the mass of Earth and sub-Earth mass planets, one would have to rely on future facilities such as Espresso/VLT (Pasquini et al. 2009) or Codex/ELT (Liske et al. 2009).  
  
We argue, however, that if the main goal is to find target planets for follow-up atmosphere observations for habitability, a planet mass is not completely necessary. This statement holds as long as the planet radius detection has a high SNR and as long as the planet is small enough. A high SNR transit detection for a small planet is sufficient to confirm planet candidacy for a transiting planet discovered from a targeted search. In contrast, a planet mass has traditionally been required for exoplanets discovered by planet transit surveys. Because transit surveys that simultaneously monitor tens of thousands of stars are fraught with false positives, a planet mass is the surest way to confirm the planetary nature of a transit signature. The argument here is that false positives are not a problem for transit signatures detected around a star with another known transiting planet, as long as the radius measurement is very robust. 

For planets with small radii in the habitable zones of their host stars, we can derive an upper mass limit, 
assuming that the most massive a planet can be is one of pure iron (Fig.~ 2). We can argue for a lower 
mass limit, using theory and models, that a small planet in the starÕs HZ will not have an H/He envelope. 
Note that an H/He envelope is usually considered bad for habitability because it traps heat making 
the planet surface too hot for complex molecules to form. Arguments about the loss of a planetÕs interior 
water reservoir (Kuchner et al. 2003; L\'eger et al. 2004) due to either stellar irradiation or from energy 
from tidal friction (e.g. Io) could be used to further theoretically constrain the planet mass.

If biosignatures or habitability features are detected, then mass estimates using future facilities are warranted, despite the huge amounts of telescope time needed. 

Our results for GJ 1214 outline the   scientific interest of the approach used by the  MEarth Project. Not only
 has the MEarth survey demonstrated its capacity to detect super-Earths transiting nearby M dwarfs, but  it also traces 
 the shortest path to the detection of habitable planets as small, or even  smaller, than the Earth and for which 
 the detection of biosignatures could be possible in the near future. In this context, we advocate the 
development of the approach used by MEarth (other facilites spread in longitude, a similar survey 
observing from Southern hemisphere, larger telescopes and IR cameras to monitor cooler M dwarfs), but also an intense
 and high-precision photometric monitoring of GJ~1214 and of the other transiting systems that MEarth (or similar projects) 
 will detect. This two-step approach targeting nearby M dwarfs permits the detection in the near-future of 
  transiting habitable planets  much smaller than our Earth which would be out  of reach for existing Doppler and transit surveys. 
   
   \begin{figure}
   \centering
   \includegraphics[width=9cm]{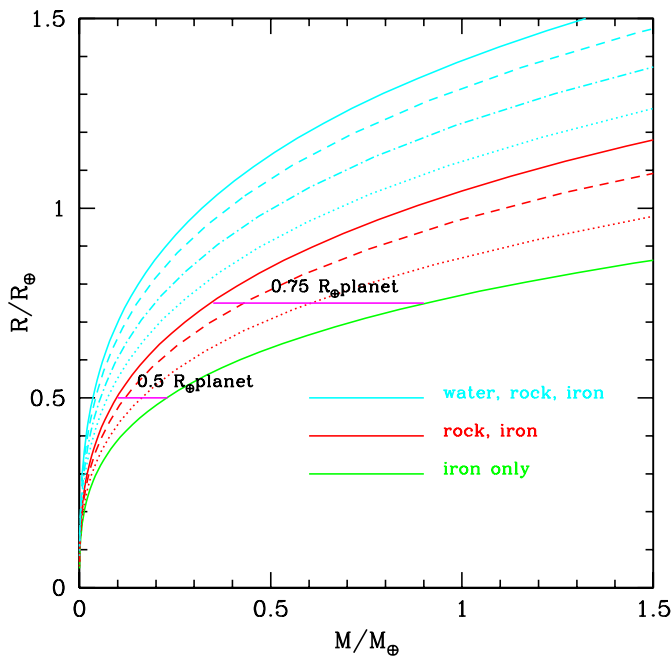}
    \caption{Mass-radius curves for exoplanets. The horizontal lines show planets of a fixed radius. An upper mass limit 
    comes from the pure iron planet case.  A lower mass limit comes from theoretical arguments, namely the inability of a 
    small and low-mass planet to retain H and He.  Further arguments about water loss may be used to argue that a planet 
    cannot have a significant interior water fraction. The solid lines are homogeneous planets. From top to bottom the 
    homogeneous planets are made of water ice (blue solid line); silicate (MgSiO3 perovskite; red solid line); and iron (Fe $\epsilon$; 
    green solid line). The non-solid lines are differentiated planets.  The red dashed line is for silicate planets with 32.5\% by 
    mass iron cores and 67.5 $\%$ silicate mantles (similar to Earth) and the red dotted line is for silicate planets with 70\% by 
    mass iron core and 30\% silicate mantles (similar to Mercury). The blue dashed line is for water planets with 75\% water ice, 
    a 22\% silicate shell and a 3\% iron core; the blue dot-dashed line is for water planets with 45\% water ice, a 48.5\% silicate 
    shell and a 6.5\% iron core (similar to Ganymede); the blue dotted line is for water planets with 25\% water ice, a 52.5\% 
    silicate shell and a 22.5\% iron core.  Curves taken from Seager et al. (2007).}
     \label{}
   \end{figure} 
      
\begin{acknowledgements}
M. Gillon is a FNRS Research Associate, and acknowledges support from the Belgian Science Policy 
Office in the form of a Return Grant.  A.~H.~M.J. Triaud researches are funded by the Swiss Fond 
National de la Recherche Scientifique. The authors thank Justin Crepp for having spotted an error in 
an equation in the first version of this manuscript. The anonymous referee is acknowledged for his 
valuable report. Last but not least, we sincerely thank NASA for believing in the idea proposed here 
and for having accepted our {\it Spitzer} GO-7 program of 485 hours of continuous observation of GJ~1214. 
\end{acknowledgements}

\end{document}